\documentclass[conference]{IEEEtran}

\usepackage{url}
\usepackage{multirow}
\usepackage{amsmath,amssymb,amsfonts}
\usepackage{algorithm,algorithmic}
\usepackage{graphicx}
\usepackage{subfigure}
\usepackage{textcomp}
\usepackage{xcolor}
\usepackage{balance}
\usepackage{cite}
\usepackage{threeparttable}

\ifCLASSINFOpdf
\else
\fi
\begin{document}
%
\title{Disappeared Command: Spoofing Attack On Automatic Speech Recognition Systems with Sound Masking}



\author{\IEEEauthorblockN{Jinghui Xu}
\IEEEauthorblockA{Tencent\\
Beijing, China\\
Email: ucasjhxu@tencent.com}
\and
\IEEEauthorblockN{Jifeng Zhu}
\IEEEauthorblockA{Tencent\\
Beijing, China\\
Email: jifengzhu@tencent.com}
\and
\IEEEauthorblockN{Yong Yang}
\IEEEauthorblockA{Tencent\\
Beijing, China\\
Email: coolcyang@tencent.com}
}

\maketitle

\begin{abstract}


The development of deep learning technology has greatly promoted the performance improvement of automatic speech recognition (ASR) technology, which has demonstrated an ability comparable to human hearing in many tasks. Voice interfaces are becoming more and more widely used as input for many applications and smart devices. However, existing research has shown that DNN is easily disturbed by slight disturbances and makes false recognition, which is extremely dangerous for intelligent voice applications controlled by voice.



The research on adversarial samples is currently mainly concentrated in the image domain. In both the physical world and the black box model, it is possible to achieve targeted adversarial attacks by slightly modifying the samples. Due to the high dimensionality of voice data and the complexity of the ASR system, it is extremely tough to implement adversarial attacks in the digital world and the physical world in the voice field. Existing black-box adversarial attack methods require frequent access to the target model to obtain evaluation scores, then adjust adversarial samples to achieve targeted attacks. Audio samples that can implement black-box attacks usually lack concealment, and the victim can easily perceive the content of the instruction. Strongly concealed adversarial attacks,  which constrain the perturbation value within a tiny range, can only achieve the attack effect on the white-box model. We propose a non-contact black-box adversarial attack algorithm with high transferability, which achieves an 81.57\% success rate of adversarial attacks on the commercially available speech API. In addition, we searched the most suitable masking music for the adversarial samples based on the psychoacoustic model to improve the concealment of the samples, and the samples after the disguise still have a 69.27\% attack success rate. We have verified the effectiveness of adversarial attacks in both the digital world and the physical world. Attackers just need ordinary speakers or mobile phones as playback devices to achieve physical adversarial attacks. The adversarial examples with masking music can attack voice applications and smart voice devices in real scenarios.

\end{abstract}

\IEEEpeerreviewmaketitle

\section{Introduction}



With the rapid development of speech recognition technology based on deep neural networks (DNN), the accuracy of the ASR system has ushered in a substantial increase, and intelligent speech services are becoming more and more popular. Products such as smart speakers and voice assistants have become part of many people's daily lives. Some online ASR services such as Google Cloud Speech-to-Text, Microsoft Bing Speech Service, Tencent Cloud Automatic Speech Recognition, and iFLYTEK voice dictation provide help for applications such as automatic meeting minutes, video analysis, and so on. ASR system enables people to interact with machines naturally and will become an important infrastructure in the era of artificial intelligence. 



AI voice services have served billions of users, and with it, the scale of the impact of its potential security threats is also increasing. Research on AI security related to adversarial attacks \cite{szegedy2013intriguing,fan2020sparse,bai2020targeted}, poisoning attacks \cite{shafahi2018poison,zhu2019transferable,xu2020approach}, backdoor attacks \cite{gu2019badnets,saha2020hidden}, etc. is constantly reminding researchers of the security issues that AI may face. Since the interpretability of DNN have not yet been resolved, most of these security issues are not easy to properly solve, and developers also lack sufficient attention to these security issues \cite{kumar2020adversarial}.


The adversarial attack is one of the most concerning artificial intelligence security problems in recent years. The DNN model can be misclassified by adding tiny perturbations to the sample that humans are difficult to perceive \cite{szegedy2013intriguing}. By adding tiny perturbations that are imperceptible to humans on the sample, DNN can be deceived and output wrong results. Current work on adversarial examples focused mainly on the image domain. Adversarial examples exist on domains, such as Natural Language Processing (NLP) \cite{jia2017adversarial} and audio, and the difficulties and challenges to be faced are also very different. This paper mainly discusses the adversarial attack of speech recognition and some work \cite{carlini2018audio} has proved that the ASR system will make wrong judgments under slight noise disturbance.


So far, although there are many mature adversarial attack algorithms in the image domain, most of them are difficult to migrate to the audio domain due to the unique characteristics of time-domain speech signals and the much more complex architecture of acoustic systems. Image adversarial samples attach tiny noises to the picture as a whole, which has almost no semantic impact on the image itself, and this disturbance is almost imperceptible to human vision. However, the adversarial samples for the ASR system must introduce additional semantics when embedding other sentences. Even if the magnitude of the disturbance is restricted as much as possible in the numerical value, it is still acoustically sensitive. In addition, when realizing an adversarial attack in the physical world, the loss of the image mainly comes from the illumination, shooting angle, and the distortion of the image carrier in the physical world. In the physical world, audio adversarial samples will not only have information loss during playback and recording but will also be disturbed by environmental noise and room reverberation.


The main challenges faced by the existing audio adversarial attack are: 1) Low success rate of transfer attacks. Currently, the most efficient black-box adversarial attack \cite{chen2020devil,zheng2021black} still requires frequent and constant access to the target model and relies on the confidence score provided by the adversary model. The non-contact black-box adversarial attack that uses the transferability of adversarial samples is not enough to generate real threats \cite{zheng2021black}. 2) Insufficient concealment. The best-hidden masking adversarial algorithm based on psychoacoustics is only effective on the white-box model and cannot be applied to black-box attacks \cite{schonherr2018adversarial,qin2019imperceptible}. The algorithms \cite{yuan2018commandersong} that can achieve black-box attacks are easy to detect the instruction information embedded in the disturbance. 3) Robustness isn't strong. Due to the loss of audio information when attacking in the real physical world, the success rate of the confrontation is much lower than the effect in the digital world \cite{zheng2021black}. The threat posed by malicious samples that can achieve adversarial attacks in the physical world is much higher than attacks in the digital world.


Most of the currently known audio adversarial attack algorithms first \cite{yuan2018commandersong,qin2019imperceptible,chen2020devil} select a piece of music or speech as the background sound and embed disturbance in it. The existence of background sound effectively masks the existence of disturbance, and it is difficult for victims to notice the specific information of disturbance under the interference of multiple sounds. On the other hand, the sound feature of the background sound similar to the target instruction can also help reduce the value of the disturbance so that uses less disturbance to realize the adversarial attack on the white-box model. In actual scenes, most of the background sounds are invalid information. These factors will have a great impact on the results of the attack when performing black-box attacks, especially when targeting business models. The adversarial attack algorithm NI-Occam without background sound proposed by \cite{zheng2021black} shows a certain degree of transferability, but due to the absence of background sound, the concealment is much worse than the previous work.


In the research of voice adversarial attacks, it is always difficult to balance the transitivity and concealment of adversarial samples. To ensure a higher success rate of transfer attacks, although the existing attack algorithm \cite{zheng2021black,chen2020devil} restricts the disturbance size during the generation process, it can still perceive obvious noise and sometimes even hear the content of the instruction. The essence of these samples to achieve the transfer attack is to allow the adversarial samples to retain the most essential audio characteristics of the target instruction with as little disturbance value as possible, so the instruction information will inevitably be easily perceived by the human ear. In the white-box model attack \cite{qin2019imperceptible}, although the perturbation can be constrained to a minimum so that the adversarial sample is 99$\%$ similar to the original sample, the generalization ability of such adversarial samples is poor, and more attacks are directed at the boundary of the model itself rather than common features. The essential characteristics of the target command are found, so it can hardly be used in black-box attacks.


In order to realize a covert contactless adversarial attack, we propose a solution that first generates a high transferability adversarial sample, and then generates the most suitable masked music for this sample. To ensure that the adversarial samples are available in the physical world, random noise is added during the generation process to improve the robustness of the samples. This solution, which takes into account transferability, concealment, and robustness, has a strong real threat and practical significance.


Generally speaking, audio black-box attacks mainly target two platforms: One is commercial Cloud Speech API, and the other is commercial devices with intelligent voice services, such as mobile phones or smart furniture. In our experiments, the attack algorithms we propose can attack these two platforms with a high success rate in both the digital world and the physical world. And in the perceptual test for embedded information, when volunteers have the authority to adjust the volume arbitrarily, they still could not clearly hear the instruction information contained in the confrontation sample with masking music added. In addition, we verified that the adversarial samples with masking music still have a certain probability of realizing adversarial attacks against commercial voice products during multiple playbacks and recordings. This is by far the most demanding verification environment, which shows that the adversarial samples we generate have transmissibility.

Our major contributions are summarized as follows:

\begin{itemize}


\item \textbf{Transferability} We propose a contactless black-box adversarial attack algorithm. Without any interaction with the target model, the generated adversarial samples can directly migrate to attack the black-box voice model with an 81.57\% attack success rate. We tested the effectiveness of adversarial examples on the commercial speech recognition APIs of iFlytek, Tencent, and Baidu.


\item \textbf{Concealment} For the first time, we proposed an optimization algorithm for background sounds, using a psychoacoustic model to search for masked music that is most suitable for hiding adversarial samples. In our test, most volunteers could not hear the information hidden in the audio clearly, and the masked samples can still be used to black-box adversarial attack with a 69.27\% attack success rate.


\item \textbf{Robustness} The adversarial samples we generate can be played in the real physical world over the air to achieve adversarial attacks. We simulated the complex scenarios that may be encountered in a real attack and used the most common household speakers and mobile phones as the playback and recording equipment. It proves that our attack has realistic threat significance.
\end{itemize}
%


In the threat model we construct, the cost of attacking by the attacker is not expensive, and the constraints are loose. Only a lightweight ASR model is needed to generate high transferability adversarial examples. Attackers can use mobile phones or radios to play the elaborate audio file or upload media files carrying malicious audio on video platforms such as YouTube to achieve adversarial attacks. The attack algorithm we proposed has strong practical significance, especially for intelligent voice devices with command functions, which may cause serious harm after being attacked. At the end of this paper, we put forward some feasible defense suggestions for the attack method.

\section{Related Work}

\subsection{Speech Recognition Systems}


The rapid development of DNN technology enables ASR systems to understand human voices with high precision, which has greatly changed the way humans and machines interact. Based on the ASR system, various companies have developed applications such as voice assistants. Smart devices that can communicate with people are gradually entering people's daily lives. People can control smart homes, mobile phones, or cars through voice. The high-performance ASR system gives developers greater confidence to open more permissions and functions for voice services. For example, a mobile phone voice assistant can use only voice to complete the entire process of downloading software or opening WeChat to send a message to a friend.

The commonly used ASR systems can be roughly divided into traditional recognition methods \cite{li2012improving,novoa2018dnn} and end-to-end recognition methods \cite{kim2021comparison,graves2006connectionist}. The main difference is reflected in the acoustic model. Traditional acoustic models generally use Hidden Markov Models (HMM) \cite{rabiner1989tutorial}, while end-to-end methods generally use DNN. After computing power is no longer a bottleneck and the training data is sufficient, the advantages of deep learning make ASR systems of different modes choose DNN as the core.

\begin{figure*}[htbp]
  \centering
  \includegraphics[width=.8\textwidth]{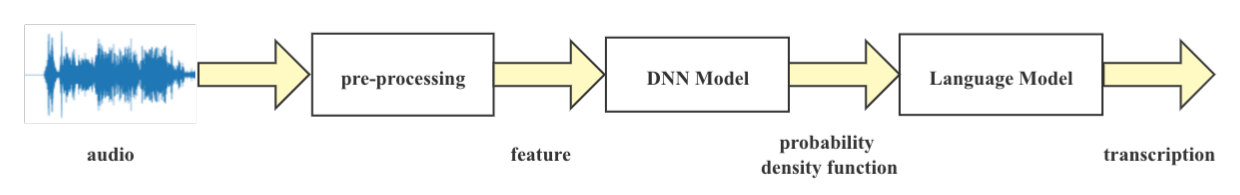} 
  \caption{The architecture of Automatic Speech Recognition System. It is mainly divided into three parts: 1) data preprocessing to extract features, 2) DNN model to return phoneme probability density distribution, 3) language model to decode and output transcription} 
  \label{asr} 
\end{figure*}

The architecture of the current speech recognition system generally includes three main steps: data preprocessing, model prediction, and decoding, as shown in Figure \ref{asr}. In the data preprocessing stage, feature extraction is performed on the received original audio data, and unnecessary information (such as information beyond the range of human hearing, noise, etc.) is discarded. Common feature extraction algorithms include Mel-frequency Cepstral Coefficients (MFCC) \cite{muda2010voice}, Linear Predictive Coefficient (LPC) \cite{itakura1975line}, Perceptual Linear Predictive (PLP) \cite{hermansky1990perceptual}, etc. The ASR system then analyzes the extracted acoustic features and predicts the most likely phoneme. In the decoding stage, the ASR system relies on the speech model to find the optimal output result according to grammatical rules, common vocabulary, and language environment. For the end-to-end model, the recognition result is output in the second step.

In order to facilitate the researchers to deploy the ASR system, some mature open source tools can be selected to help build the white-box model. Kaldi is an open-source speech recognition toolkit written in C++ and is available for free under Apache License v2.0. Kaldi aims to provide flexible and scalable components, including a variety of speech signal processing, speech recognition, voiceprint recognition, and DNN. Kaldi has been widely welcomed by researchers, and we built a language model for generating adversarial examples based on Kaldi. \footnote{https://github.com/kaldi-asr/kaldi}

\subsection{Adversarial Examples}


The research in this field mainly originated from \cite{biggio2013evasion,szegedy2013intriguing}, by adding a small perturbation model that is imperceptible to the human eye to make untargeted wrong judgments on pictures. In order to explore the potential hazards of adversarial samples more comprehensively, researchers in the field of Computer Vision (CV) have taken the lead in continuous in-depth research and has made progress in targeted attacks \cite{chen2020boosting,feng2021boosting}, transfer attacks \cite{huang2019black}, and physical attacks \cite{eykholt2018robust,sharif2016accessorize}. These works proved the fragility of neural networks and inspired researchers in other fields of deep learning to explore similar vulnerabilities.


In the field of NLP, the data used is all discrete data. The definition of tiny disturbances is different from the CV field. Words with similar meanings are usually used to replace original words to reduce the degree of human perception of disturbances. Therefore, the CV field adversarial attack algorithms usually cannot be directly transformed to use. Although the data types are quite different, the methods in the CV field still have a lot of reference significance. Especially the use of gradient information has important reference significance in adversarial attacks in different fields.


Adversarial attacks are more difficult to advance in the field of audio. Early works \cite{gong2017crafting,cisse2017houdini} and others have implemented untargeted adversarial attacks on DNN-based ASR systems. Commandersong \cite{yuan2018commandersong} first achieved targeted adversarial attacks while ensuring invisibility by introducing additional background music. In the same year, in order to better hide the adversarial samples, \cite{schonherr2018adversarial} proposed to use a psychoacoustic model to hide adversarial perturbation. Before this, the concealment of adversarial attacks was mainly ensured by constraining the magnitude of the disturbance value, which is effective on images, but the voice adversarial samples need to continue to carry additional information for a while, and it is almost impossible to control the value of the disturbance within a range that humans cannot perceive. In most cases, the victim can feel an obvious sense of noise and have a certain chance to guess the hidden information. \cite{schonherr2018adversarial} inspired us to enhance the concealment of samples only to deceive human senses, and it is not necessary to restrict the value of perturbation.


Adversarial algorithms against DNN defects are not the only means to achieve concealed attacks. There are many ways to shield the perception of human hearing to protect hidden information. The dolphin attack \cite{zhang2017dolphinattack} uses an ultrasonic modulation command that is not perceptible to the human ear to achieve a "silent" control voice system. However, this type of ultrasonic attack has a short distance and requires special playback equipment, which makes it very difficult to achieve a real attack. Enhancing playback equipment \cite{roy2018inaudible} or using solid-state sound \cite{yan2020surfingattack} is a way to increase the tool distance, but these attacks require harsh implementation conditions. 


The entry point of another concealment adversarial attack is at the feature extraction stage. MFCC is often used for the conversion of audio features. It can effectively reduce the dimensionality of data. It is widely used in both traditional ASR systems and deep learning-based ASR systems. MFCC features are many-to-one and irreversible. Therefore, the attacker can find segments that are consistent with the MFCC features of the target command but have completely different audio information to achieve an adversarial attack against the MFCC algorithm \cite{vaidya2015cocaine,carlini2016hidden,abdullah2019practical,abdullah2021hear}. The audio contained in the audio produced by this type of attack method is human. What the senses cannot understand.


To be able to achieve adversarial attack in the real physical world, the image field simulates the possible changes (distortion, rotation, etc.) of the printed picture in the physical world during the process of generating adversarial samples to achieve physical attack \cite{wu2020making}. Audio adversarial attack will encounter more environmental factors in the physical world. In order to improve the success rate of attacks in the physical world, Yakura et al. \cite{yakura2018robust} proposed to simulate the conversion (such as reverberation and noise) caused by playback or recording in the physical world and incorporate the conversion into the generation process to obtain a robust adversarial sample. By simulating a more complex real environment, it is possible to achieve target attacks in different physical world scenarios \cite{li2020advpulse,chen2020metamorph}.


The work of (2020 Yuxuan Chen) on the basis of previous work, by relaxing the limit of the disturbance value and multiple visits to the target model, proved for the first time that the migration attack of speech recognition is feasible in both the digital world and the physical world.

The previous work was partially unified by \cite{qin2019imperceptible}, using psychoacoustic models and gradient descent, the white box model achieved a highly concealed, 100$\%$ accurate targeted attack. Humans can hardly hear the extra commands embedded into the audio. This proves that the perturbation in speech can be constrained to a minimum on the white-box model, but this method cannot be applied to black-box attacks. The work of \cite{chen2020devil} on the basis of previous work, by relaxing the limit of the disturbance value and multiple visits to the target model, proved for the first time that the transfer attack of speech recognition is feasible in both the digital world and the physical world. Furthermore, \cite{zheng2021black} implements two black-box attack methods: query and contactless.


Most of them choose fixed background audio to hide the adversarial samples, which makes the adversarial samples not necessarily generated on the most suitable location. Therefore, we mainly refer to the work of \cite{schonherr2018adversarial,qin2019imperceptible} using the psychoacoustic model to hide the adversarial sample and find the least syllable to hide the generated adversarial sample.

\section{Background}

\subsection{Problem Definition}


For a given input audio $x$ (maybe a song or a sentence) and the corresponding target transcription $y$, a DNN-based ASR system can be simply regarded as a mapping function $f(\cdot)$, and satisfies $f(x)=y$. In fact, there are many complicated parameters in $f(\cdot)$, and our white-box model includes the process of predicting the phoneme of each frame from the speech feature and then converting it into text.


Our object is to construct an imperceptible and targeted adversarial example $x'$ that can attack the ASR system when played over the air. The usual method of adversarial attack is to fix the audio $x$ and seek a small disturbance $\delta$, let $x' = x + \delta, f(x)=y , f(x')=y', (y \neq y')$, and satisfy $x\prime$ sounds similar to $x$ so that the victim is not easily aware of the disturbance.


The purpose of the adversarial attack is to hide additional information in the audio and make it difficult for people to perceive. Due to the high dimensionality of voice data and the complexity of the ASR system, the threshold $\delta$ when searching for a voice adversarial sample disturbance is much larger than that in the image. Therefore, when the adversarial attack algorithm in the image domain is transferred to the audio domain, it often does not have enough concealment. In order to solve the problem of concealment, \cite{schonherr2018adversarial,qin2019imperceptible} 
take advantage of the temporal masking and frequency masking characteristic of human hearing, and use a psychoacoustic model to constrain the value of the disturbance. This ensures that even if $\delta$ is relatively large in value, it is still difficult for people to hear it.


However, this method does not guarantee that the background sound used to hide the disturbance is appropriate. Different from previous method, in our work, we first use the gradient descent algorithm to find a $\delta$ as small as possible, let $f(\delta)=y'$, and then find $x$ suitable for $\delta$ ($x$ can be an audio segment composed of several notes), satisfying $f(x + \delta)=y'$, and $x + \delta$ is difficult for the human ear to perceive $y'$.

\subsection{ASR model}


We adopted the DataTang TDNN Chain Model in Kaldi \cite{povey2011kaldi,wang2019datatang} as the white-box model, which was trained on 1505 Chinese Mandarin corpus released by DataTang. Kaldi is currently one of the most popular open-source speech toolkits. Kaldi is also used in some commercial products to process audio input \cite{schonherr2018adversarial}. In terms of data set size and model structure, our Chinese language model is a lightweight model, and can still successfully attack commercial speech recognition models trained on large corpus data sets.


The ASR system we built is shown in Figure \ref{asr}. In the feature extraction stage, we use the MFCC algorithm to extract audio features, which is currently one of the most popular methods. This process converts the original time-domain signal into a frequency domain signal and obtains a spectrogram of the audio signal, then inputs it into the DNN model. DNN will calculate the phonemes of each frame with preset steps. 

The output of DNN is the probability density function (pdf) \cite{yuan2018commandersong}. The p.d.f. is indexed by the pdf identifier (pdf-id), which exactly indicates the column of the output matrix of DNN. Kaldi can decode the pdf-id sequence to convert the phoneme sequence into a sentence.

\subsection{Adversarial Sample}

In this part, we introduce the generation method of audio adversarial samples. Generally speaking
, given an audio sample $x$ and a target phrase $y$, we formulate the problem of constructing an imperceptible adversarial example $x' = x + \delta $ as minimizing the loss function  $\ell (x, \delta, y)$, which is defined as:

\begin{equation}
\ell (x,\delta ,y) = \ell_{net}(f(x+\delta ),y)+\alpha \cdot \ell_{\theta }(x,\delta ) \label{equ:ad}
\end{equation}

where $\ell_{net}$ requires that the adversarial examples fool the audio recognition system into making a targeted prediction $y$, where $f(x) \neq y$. $\ell_{\theta}$ is used to constrain the size of the disturbance to reduce the degree of human hearing perception of audio. $\alpha$ is used as the perception coefficient to control the amplitude of the disturbance, and the value of $\alpha$ is set according to prior knowledge.

We encountered a similar problem with \cite{qin2019imperceptible}. Under such a setting, the model usually chooses one of the losses for optimization or neither, which may be caused by too large a search space. Therefore, we refer to the \textbf{two stage attack} of \cite{qin2019imperceptible}. The first stage restricts the disturbance to generate a phoneme sequence consistent with the target sentence, and the second stage reduces the size of the disturbance itself.

\subsection{Two stage attack}

In the first stage, we let the adversarial samples match the phonemes of the target sentence. The generation method of this stage refers to the pdf-id matching algorithm proposed in CommanderSong\cite{yuan2018commandersong} to generate adversarial samples. This method has been proved in CommanderSong\cite{yuan2018commandersong} and Devil's whisper\cite{chen2020devil} to achieve transfer attacks, and it can still achieve adversarial attacks when played over-the-air.

we set $\alpha$ in Eq.(\ref{equ:ad}) to be zero and clip the perturbation to be within a relatively small range (This threshold is also set based on prior knowledge and is not too small). As a result, the first stage solves:

\begin{equation}
\begin{aligned}
\mathbf{minimize} \quad &\ell_{net}(f(x+\delta),y) \\
\mathbf{such \ that}\quad &\left \| \delta \right \| < \epsilon 
\end{aligned}
\end{equation}

In order to improve the robustness and enable it to attack in the physical world, we refer to CommanderSong\cite{yuan2018commandersong} and Devil’s whisper\cite{chen2020devil} and add additional noise to $x$ during the process of generating adversarial samples, that is, the Eq.(\ref{equ:ad}) becomes:

\begin{equation}
\ell_{net}(f(x+\delta),y) \rightarrow \ell_{net}(f(x+\delta+z)),z \sim N(0,\sigma ^2)
\end{equation}

Therefore, the loss becomes:

\begin{equation}
\ell (x,\delta ,y) = \ell_{net}(f(x+\delta+z),y)+\alpha \cdot \ell_{\theta }(x,\delta ) \label{equ:ad2}
\end{equation}

where $\left \| \delta \right \|$ represents the $\left \| \cdot \right \|_{\infty}$ max-norm of $\delta$. Specifically, we set $\delta = 0$ when initialization and then on each iteration:

\begin{equation}
\delta \leftarrow clip_{\epsilon}(\delta - lr \cdot sign(\nabla_{\delta} \ell_{net}(f(x+\delta +z),y))),\label{equ:stage1}
\end{equation}

where $lr$ is the learning rate and $\nabla_{\delta} \ell_{net}$ is the gradient of $\ell_{net}$ with respect to $\delta$. We initially set $\epsilon$ to a large value. Unlike the work of \cite{carlini2018audio}, we did not gradually reduce the value of $\epsilon$ during optimization, and the looser disturbance constraints can complete the first step faster.

The second stage focuses on making the adversarial examples imperceptible with an unbounded max-norm. in this stage, $\delta$ is mainly constrained by the size of the disturbance. Specifically, initialize $\delta$ with $\delta_{im}^{*}$ optimized in the first stage and then on each iteration:

\begin{equation}
\delta \leftarrow \delta - lr \cdot \nabla_{\delta} (\ell(x,\delta,y) + \alpha \cdot \ell_{\theta }(x,\delta )), \label{equ:stage2}
\end{equation}


where $lr$ is the learning rate inherited from the first stage and $\nabla_{\delta}\ell$ is the gradient of $\ell$ with respect to $\delta$. The loss function is defined in Eq.(\ref{equ:ad}). The parameter $\alpha$ that balances the network loss $\ell_{net}(f(x+\delta+z))$ and the imperceptibility loss $\ell_{\theta} (x,\delta)$ is initialized with a small value, e.g., 0.001.


Our entire process of generating non-contact black-box attack adversarial samples is shown in Algorithm \ref{alg:ad}. The algorithm is basically similar in principle to the NI-Occam algorithm proposed by \cite{zheng2021black}. By restoring the key parts of the natural command audio, adversarial samples will achieve a high success rate of transfer attack. This method does not require any interaction with the target model to achieve transfer attacks, and the ability of black-box attacks is very strong. But the disadvantage is that the generated samples have very poor concealment and can perceive the natural commands contained in the disturbance.

\begin{algorithm}
	\renewcommand{\algorithmicrequire}{\textbf{Input:}}
	\renewcommand{\algorithmicensure}{\textbf{Output:}}
	\caption{Algorithm for generating music hiding adversarial samples based on heuristic search}
	\label{alg:ad}
	\begin{algorithmic}[1]
	
		\REQUIRE ASR model $f$, target sentense $y$, perception coefficient $v_{\alpha}$, Loop limit $I$, learning rate $l_r$
		\ENSURE adversarial sample $\delta$
        \STATE Initialize $\delta = 0 \ \alpha = 0$
        \STATE Gets the target phoneme $y_p = M(y)$
        \WHILE{not converged yet}{
            \STATE Sample $z\sim \mathcal{N} (0,\sigma ^2)$.
            \STATE Use the Optimizer to minimize the loss in Eq. (\ref{equ:ad2}) with $l_r$ and update $\delta$.
            \STATE Clip $\delta$ into the constrained range.
            \IF {$f(\delta)=f(y)$}
                \STATE Update $\alpha = v_{\alpha}$ in Eq. (\ref{equ:ad2})
            \ENDIF 
            }
        \ENDWHILE
    \STATE \textbf{return} $\delta$
	\end{algorithmic}
\end{algorithm}


Different from \cite{qin2019imperceptible}, we don't use background sound for masking in the generation process, and naturally there will be no corresponding loss items based on the psychoacoustic model. In the initialization phase, our algorithm is similar to \cite{zheng2021black}, starting from the silent segment to optimize $\delta$.

\section{Attack Method}\label{AA}


Unlike on images, where minimizing $\ell_{p}$ distortion between an image and the nearest misclassified example yields a visually indistinguishable image, on audio, this is not the case \cite{schonherr2018adversarial,qin2019imperceptible}. Thus, in this work, we depart from the $\ell_{p}$ distortion measures and instead rely on the extensive work which has been done in the audio space for capturing the human perceptibility of audio.

\subsection{Psychoacoustic Models}\label{psm}

A good understanding of the human auditory system is critical in order to be able to construct imperceptible adversarial examples. In this paper, we take advantage of temporal masking and frequency masking to hide adversarial samples. Simply put, the mask can be seen as creating a "mask threshold" in the frequency domain. Any signal below this threshold is practically imperceptible.

\begin{figure}
  \centering
  \includegraphics[width=.45\textwidth]{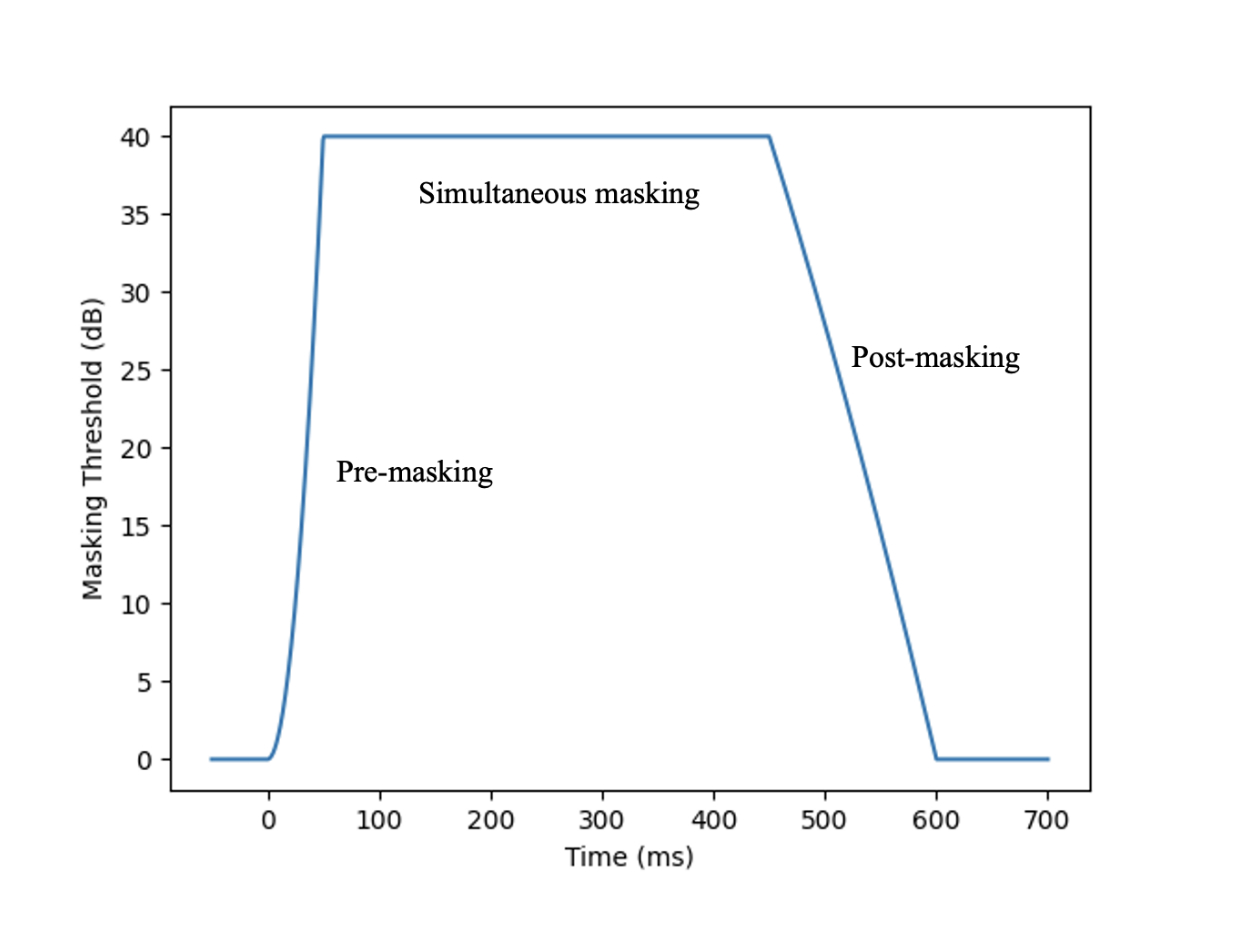} 
  \caption{Schematic diagram of the time masking effect. According to the time of sound existence in the time domain, it is divided into pre-masking, simultaneous masking, and post-masking. The advanced masking is only effective for a very short time, that is, 20ms. The post-masking strength decays exponentially with time until it becomes 0 after 100-200 ms.} 
  \label{fig:time} 
\end{figure}

The time-domain masking effect \cite{fastl2007information,zwicker2013psychoacoustics} is the masking effect between adjacent sounds in the time domain, or it is called non-simultaneous masking. Time-domain masking is divided into pre-masking or backward masking and post-masking or forward masking. When the masked sound and the masked sound exist at the same time, simultaneous masking occurs, that is, frequency-domain masking \cite{bosi2002introduction,lin2015principles}. A simple understanding of time-domain masking is shown in Figure \ref{fig:time}. Audio signals that are below the threshold volume of the masking sound at the corresponding time are difficult to perceive.

\begin{figure}
  \centering
  \includegraphics[width=.45\textwidth]{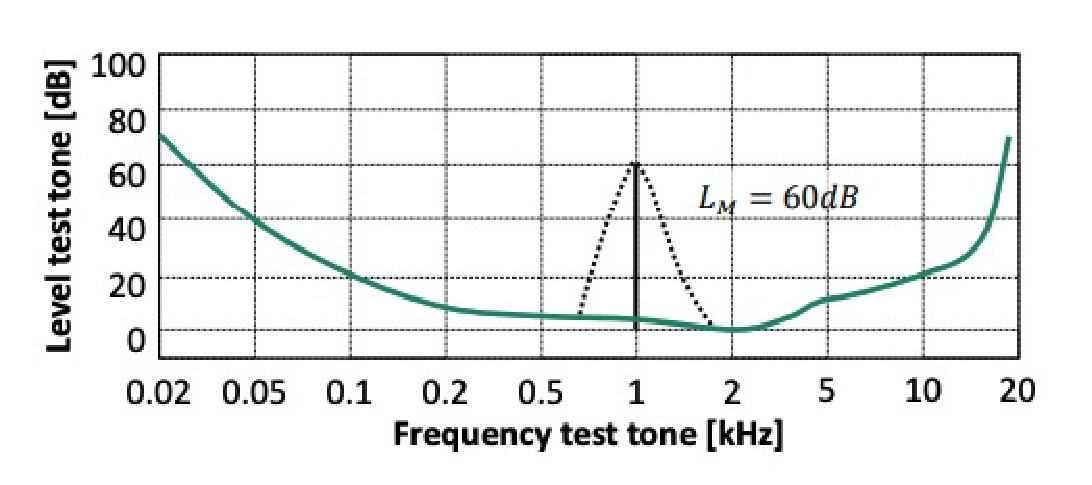} 
  \caption{Schematic diagram of frequency-domain masking effect\cite{schonherr2018adversarial}. The hearing threshold of the test tone (dotted line) is masked by an $L_M=60dB$ tone at 1 kHz \cite{zwicker2013psychoacoustics}. Green indicates the hearing threshold in a quiet state.} 
  \label{fig:fre} 
\end{figure}

When the masking sound and the masked sound exist at the same time, simultaneous masking occurs, that is, frequency-domain masking, as shown in Figure \ref{fig:fre}. A strong pure tone will mask the weak pure tones that sound at the same time near its frequency. For example, a pure tone with a sound intensity of 32 dB and a frequency of 850 Hz will be masked by a pure tone with a sound intensity of 50 dB and a frequency of 840 Hz. Because the sound frequency and the masking curve are not linear, in order to measure the sound frequency uniformly from the perception, the Bark scale is introduced for measurement \cite{zwicker1961subdivision}.

We basically refer to \cite{qin2019imperceptible} to calculate the masking threshold based on the psychoacoustic model and guide the search for masking music.

\subsubsection{step1: Identifications of Maskers}\label{iom}

In order to calculate the masking threshold, we first compute the short-time Fourier transform of the raw audio signal to obtain the spectrum of overlapping sections (called “windows”) of a signal. The window size $N$ is 2048 samples which are extracted with a "hop size" of 512 samples and are windowed with the modified Hann window. We denote $s_{x}(k)$ as the $k$th bin of the spectrum of frame $x$.

Then, we compute the log-magnitude power spectral density (PSD) as follows:

\begin{equation}
p_{x}(k) = 10\mathrm{log} _{10}\left |\frac{1}{N} s_{x}(k)  \right |^{2}
\end{equation}

The normalized PSD estimate $\bar{p}_x (k)$ is defined by \cite{lin2015principles}:

\begin{equation}
\bar{p}_{x}(k)=96-\underset{k}{\mathrm{max}} \left \{ {p_{x}(k)} \right \} +p_{x}(k)
\end{equation}

Given an audio input, in order to compute its masking threshold, first we need to identify the maskers, whose normalized PSD estimate $\bar{p}_{x}(k)$ must satisfy three criteria. First, they must be local maxima in the spectrum, satisfying:

\begin{equation}
\bar{p}_{x}(k-1)\leqslant \bar{p}_{x}^{m}(k) \quad and \quad \bar{p}_{x}^{m}(k) \geqslant \bar{p}_{x}(k+1)
\end{equation}

where $0\leqslant k\leqslant \frac{2}{N}$.

Second, the normalized PSD estimate of any masker must be higher than the threshold in quiet $ATH(k)$, which is:

\begin{equation}
\bar{p}_{x}^{m}(k) \geqslant \mathrm{ATH}(k),
\end{equation}

where $ATH(k)$ is approximated by the following frequencydependency function:

\begin{equation}
\begin{split}
\mathrm{ATH} (f) &= 3.64(\frac{f}{1000})^{-0.8}-6.5\mathrm{exp} {-0.6(\frac{f}{1000}-3.3)^2}\\
& +10^{-3}(\frac{f}{1000})^4
\end{split}
\end{equation}

The quiet threshold only applies to the human hearing range of $20\mathrm{Hz} \leqslant f\leqslant 20\mathrm{kHz}$. When we perform short time Fourier transform (STFT) to a signal, the relation between
the frequency $f$ and the index of sampling points $k$ is

\begin{equation}
f=\frac{k}{N} \cdot f_{s}, \quad 0 \leqslant f \leqslant \frac{f_{s}}{2}
\end{equation}

where $f_{s}$ is the sampling frequency and $N$ is the window size.

Last, the maskers must have the highest PSD within the range of $0.5$ Bark around the masker’s frequency, where bark is a psychoacoustically-motivated frequency scale. Human’s main hearing range between $20\mathrm{Hz}$ and $16\mathrm{kHz}$ is divided into $24$ non-overlapping critical bands, whose unit is Bark, varying as a function of frequency $f$ as follows:

\begin{equation}
b(f)=13\mathrm{arctan} (\frac{0.76f}{1000}) + 3.5 \mathrm{arctan} (\frac{f}{7500}) ^2
\end{equation}

As the effect of masking is additive in the logarithmic domain, the PSD estimate of the the masker is further smoothed with its neighbors by:

\begin{equation}
\bar{p}_x^m (\bar{k} ) = 10\mathrm{log} _{10}\left [ 10^{\frac{\bar{p}_x (k-1) }{10}} + 10^{\frac{\bar{p}_x^m (k) }{10}} + 10^{\frac{\bar{p}_x (k+1) }{10}} \right ]
\end{equation}

\subsubsection{step2: Individual masking thresholds}\label{imt}

An individual masking threshold is better computed with frequency denoted at the Bark scale because the spreading functions of the masker would be similar at different Barks. We use $b_{(i)}$ to represent the bark scale of the frequency index $i$. There are several spreading functions introduced to imitate the characteristics of maskers and here we choose the simple two-slope spread function:

\begin{equation}
\mathrm{SF} \left [ b_{(i)},b_{(j)} \right ] = \left\{\begin{matrix}
27\Delta b_{ij} \le 0.\\
G(b(i)) \cdot \Delta b_{ij},  \quad \mathrm{othewise} \end{matrix}\right.
\end{equation}

where:

\begin{equation}
\Delta b_{ij} = b(j) - b(i)
\end{equation}

\begin{equation}
G(b(i)) = \left [ -27 + 0.37 \mathrm{max}\left \{ \bar{p}_x^m (b(i))-40,0 \right \}  \right ] 
\end{equation}

where $b(i)$ and $b(j)$ are the bark scale of the masker at the frequency index $i$ and the maskee at frequency index $j$ respectively. Then, $T\left [ b(i),b(j) \right ]$ refers to the masker at Bark index $b(i)$ contributing to the masking effect on the maskee at bark index $b(j)$. Empirically, the threshold $T\left [ b(i),b(j) \right ]$ is calculated by

\begin{equation}
T\left [ b(i),b(j) \right ] =\bar{p}_x^m (b(i)) + \Delta_m \left [ b(i) \right ] +\mathrm{SF} \left [ b(i),b(j) \right ] 
\end{equation}

where $\Delta_m \left [ b(i) \right ] = -6.025 - 0.275b(i)$ and $\mathrm{SF} \left [ b(i), b(j) \right ]$ is the spreading function.

\subsubsection{step3: Global masking threshold}\label{gmt} 

The global masking threshold is a combination of individual masking thresholds as well as the threshold in quiet via addition. The global masking threshold at frequency index $i$ measured with Decibels (dB) is calculated according to:

\begin{equation}
\theta_x (i) = 10 \mathrm{log}_{10} \left | 10^{\frac{ATH(i)}{10} } + \sum_{j=1}^{N_m} 10^{\frac{T\left [ b(j),b(i) \right ] }{10} } \right | 
\end{equation}

where $N_m$ is the number of all the selected maskers. The computed $\theta_x$ is used as the frequency masking threshold for the input audio $x$ to construct imperceptible adversarial examples.

The code for the realization of the psychoacoustic masking effect can be founded in CleverHans (a Python library to benchmark machine learning systems' vulnerability to adversarial examples ) \cite{papernot2018cleverhans} \footnote{https://github.com/yaq007/cleverhans/tree/master/examples/adversarial\_asr}.

\subsection{Search Masking Music}\label{smm} 



Taking the masking threshold as the evaluation criterion of the masking effect, we can find the most suitable masking music by constantly adjusting the music. Therefore, heuristic search is a very suitable method.


After completing the calculation of the masking threshold, we need to search for new and better music and generate a new $x$. For the search-style masking music generation method, we try different combinations of notes to find the most suitable masking music. We need to collect music clips of different timbres $\gamma_{m,n,p,q}$, where index $m$ means different tones, $n$ means different timbre, $p$ means different duration, and $q$ means different position.

For the number $k$ of music pieces $\gamma$, we set it according to prior knowledge, that is, set the value of $k$ when there are several important syllables. For example, as for "ok" audio, k will be set $2$.

Our heuristic search method is shown in Algorithm \ref{alg:search}.

\begin{algorithm}
	\renewcommand{\algorithmicrequire}{\textbf{Input:}}
	\renewcommand{\algorithmicensure}{\textbf{Output:}}
	\caption{Algorithm for generating music hiding adversarial samples based on heuristic search}
	\label{alg:search}
	\begin{algorithmic}[1]
		\REQUIRE adversarial sample $\delta$, silence audio $x_s$, frame length $l$, different music footage $\gamma$, Loop limit $I$, ASR model $M$, target label $y$
		\ENSURE music hiding adversarial samples $s_{best}$
        \STATE Divide $s$ according to $l$, and get the set of positions $L$ where $\gamma$ can be inserted.
        \STATE Compute $bark$ and $ATH$ base on gongshi1
        \STATE Initial $\gamma1, \gamma1 \cdots \gamma_k$
        \STATE $s_t = s + \gamma_1 + \cdots + \gamma_k$
        \STATE $s_{best},v_{best}=s_t, +\infty $
        
        \FOR {$i=0$ to $I$}
            \STATE Compute $psd_t$ and $psd_{max}$ of $s_t$ base on Section \ref{iom}
            \STATE Compute $theta$ of $s_t$ with $bark,ATH,psd_t$ base on Section \ref{imt}
            \STATE Compute $theta_{\delta}$ with $psd_{max}$ base on Section \ref{gmt}
            \STATE Compute $v_t = \mathrm{mean} ( \mathrm{abs}(theta_{\delta} - theta) )$
            \IF {$v_t < v_{best} \ \mathrm{and} \ M(s_t) = y$}
                \STATE Update $s_{best},v_{best} = s_t,v_t$
            \ENDIF 
        \ENDFOR
		\STATE \textbf{return} $s_{best}$
	\end{algorithmic}
\end{algorithm}


Since the length of the voice data depends on the sampling rate and duration, in order to speed up the search, we divide the audio into frames in Algorithm \ref{alg:search}. When inserting $\gamma$, it is only executed at the beginning of each frame, which means the best position that may be missed when searching, but this is acceptable within the error tolerance. We recommend setting the frame length not to exceed 200 ms (this is about the upper limit of the decay period of the post-masking effect). After completing the search, we use the FFmpeg tool to synthesize the masking music and the adversarial samples to obtain the final attack samples.

\section{Evaluation}

\textbf{Evaluation Metrics}


In order to evaluate the efficiency of adversarial attacks and the degree of influence of masking music on attacks, we measure the success rate of attack (SRoA) before and after the masking music is added. Generally, we use the Chinese-character-level success rate of attack $\mathrm{SRoA_c}=\frac{c}{N_c}$ for evaluation. $c$ is the number of Chinese characters consistent with the target sentence and $N_c$ is the number of Chinese characters in the adversarial samples.

Since there are many Chinese characters with similar pronunciations, so as to evaluate the masking effect more fine-grained, we also use pinyin-level (Chinese characters conversion to Chinese Pinyin) success rate of the attack $\mathrm{SRoA_p}=\frac{p}{N_p}$ to measure. $p$ is the number of pinyin characters consistent with the target sentence and $N_p$ is the number of pinyin characters in the adversarial samples.

Because we are not embedding additional target instructions through perturbation in the existing audio information, the commonly used measure of SNR (the ratio of the signal power of the original audio to the noise power) \cite{chen2020devil,zheng2021black} is not suitable for evaluating our attack algorithms.


As for the evaluation of the masking effect on the human senses, because the evaluation is subjective, we interviewed several volunteers (the interviewed volunteers have been anonymously processed) to measure the masking effect based on their intuitive feelings. We use two indicators, Once-recognize Rate and Twice-recognition rate, to evaluate the recognition accuracy of volunteers, that is, the probability that volunteers can accurately recognize words during the first listening and the second listening.

\textbf{Experiment Setting}


In order to evaluate the attack effect and physical realization effect of adversarial samples, we designed two sets of experiments: 1) Adversarial attacks on publicly available commercial speech API services. 2) Play and record adversarial samples in the real physical world to attack and test them on commercial speech APIs.


We chose the commercial APIs of iFlytek, Tencent, and Baidu for testing. They provide a wide range of Chinese voice system services and affect the user experience of hundreds of millions of people. Attacks on this type of API are closer to the real attack scenario and have higher difficulty. When selecting the target label, we selected some functional instructions as the attack target \footnote{target command includes: play music, restart, open document, close computer, shut down, browser, search, increase brightness, WeChat, enter, etc.}. In this setting, a high success rate means a wider range of potential risks.


In terms of physical attack experiments, we simulated real attack scenarios in an ordinary bedroom environment for experiments. We use Sony's SRS-XB01 speaker, an ordinary home Bluetooth speaker, as a playback device. Use Huawei Honor 30 mobile phone as a recording device. The distance between the speaker and the target device is about 15cm. A cluttered environment with more furniture will bring more complex echoes, and ordinary home audio and mobile phones as audio playback and recording equipment will bring more losses when performing physical attacks. The experimental evaluation environment we set up is universal and demanding, completely simulating real and complex attack scenarios.


When preparing the material for the masking music, in order to reduce the impact of the quality of the music material, we use Mido (a library for working with MIDI messages and ports) \footnote{https://github.com/mido/mido} to generate Musical Instrument Digital Interface (midi) files \cite{smith1981the} composed of different tones and timbres, and convert them into .wav files. The MIDI protocol stipulates a standard MIDI musical instrument timbre sorting table, which contains 16 commonly used musical instruments, each with 8 timbres, and a total of 128 timbres. Audio files converted from midi files will not be affected by any environmental factors, loss during playback and recording, etc., and the impact on adversarial samples can be reduced as much as possible. 

\subsection{Evaluation on Cloud Speech APIs}




Table \ref{tab:digital} records the targeted attack performance of the adversarial samples we generated on different commercial ASR system service APIs. Without adding hidden sounds, the success rate of word-level adversarial attacks on different commercial APIs is between 74\% and 88\%, while the success rate of pinyin-level adversarial attacks is higher, between 83\% and 92\%. The average values of $SRoA_c$ and $SRoA_p$ are 81.57\% and 88.1\%, respectively.

\begin{table}[htbp]
\begin{center}
\centering
\caption{Experimental results on targeted attacks on commercial cloud speech-to-text APIs}
\begin{tabular}{|c|cc|cc|}
\hline
                       & \multicolumn{2}{c|}{Without Masking Music}     & \multicolumn{2}{c|}{With Masking Music}        \\ \hline
\multicolumn{1}{|c|}{} & \multicolumn{1}{c|}{$\mathrm{SRoA_c}$}  & $\mathrm{SRoA_p}$  & \multicolumn{1}{c|}{$\mathrm{SRoA_c}$}  & $\mathrm{SRoA_p}$  \\ \hline
iFLYTEK                & \multicolumn{1}{c|}{74.0}  & 83.65 & \multicolumn{1}{c|}{63.45} & 70.74 \\ \hline
Tencent API            & \multicolumn{1}{c|}{88.37} & 92.91 & \multicolumn{1}{c|}{76.4}  & 84.25 \\ \hline
Baidu API              & \multicolumn{1}{c|}{82.35} & 87.73 & \multicolumn{1}{c|}{67.95} & 76.15 \\ \hline
\end{tabular}
\label{tab:digital}
\end{center}
\end{table}

Compared with the 52\% attack success rate of black-box non-contact attacks \cite{zheng2021black}, it has been greatly improved, and it is currently the most successful adversarial attack method for migration attacks. The success rate of the Pinyin-level confrontational attack is slightly higher than that of the word-level. This is because some words are very similar in pinyin, but the similarity is lower at the character level, such as the two words "hui che"("enter" in English) and "huo che"("train" in English). The Chinese-character-level difference is 50\%, but only one character difference in pinyin-level. Of course, the high attack success rate of the adversarial sample is brought by sacrificing invisibility, although the content it contains does not sound very clear.


After the addition of masking music, the success rate of the adversarial attack dropped by more than 10\% at the character-level compared with the sample in the silent background, and the success rate at the pinyin-level decreased slightly. This is mainly due to the influence of the masking sound on the recognition of some phonemes. A small change in pinyin will cause the recognition result to become homophones, near-phonetic words with similar pronunciation. From the perspective of the success rate of adversarial attacks, the masking sound does not have much influence on the success rate of adversarial attacks, and it is an effective way of masking.

\subsection{Evaluation on Physical World}\label{sec:phy}


We have also tested the effectiveness of adversarial samples and masking sound effects in the physical world. The adversarial attack in the physical world is far more complicated than the digital world. During the experiment, we chose a natural bedroom (not pursuing an extremely quiet and unobstructed laboratory environment) as the experimental site and controlled the playback equipment, recording equipment, playback distance, volume, and other factors to not change. Interestingly, the same sample may produce different results during repeated playback, and in most cases, these results are extremely similar. These uncertain factors are one of the biggest challenges to be faced in physical attacks. 


In order to facilitate multiple recordings for experiments, we use speakers to play the adversarial samples every 2 seconds for continuous recording. The divided samples will be tested on the speech API and compared with the experimental results in Table \ref{tab:digital}. Our experiment fully considers the factors that may affect the results such as playback equipment, recording equipment, room reverberation, and natural noise in a real physical adversarial attack. The statistical results of the physical attack experiment are shown in Table \ref{tab:physical},.


\begin{table}[htbp]
\begin{center}
\centering
\caption{Experimental results on targeted attacks on commercial cloud speech-to-text APIs in the physical world}
\begin{tabular}{|c|cc|cc|}
\hline
                       & \multicolumn{2}{c|}{Without Masked Music}     & \multicolumn{2}{c|}{With Masked Music}        \\ \hline
\multicolumn{1}{|c|}{} & \multicolumn{1}{c|}{$\mathrm{SRoA_c}$}  & $\mathrm{SRoA_p}$  & \multicolumn{1}{c|}{$\mathrm{SRoA_c}$}  & $\mathrm{SRoA_p}$  \\ \hline
iFLYTEK                & \multicolumn{1}{c|}{67.6}  & 78.8 & \multicolumn{1}{c|}{59.64} & 65.79 \\ \hline
Tencent API            & \multicolumn{1}{c|}{83.96} &90.47 & \multicolumn{1}{c|}{64.95}  & 75.82 \\ \hline
Baidu API              & \multicolumn{1}{c|}{73.29} & 79.93 & \multicolumn{1}{c|}{57.32} & 64.28 \\ \hline
\end{tabular}
\label{tab:physical}
\end{center}
\end{table}

Compared with Table \ref{tab:digital}, the success rate of adversarial attacks in the physical world has a certain decrease in different modes. The average success rate of character-level without masking sound effects is 74.95\%. After the masking sound effect is added, the average success rate is 60.64\%. Compared with the previous black box confrontation attacks, the two methods have greatly improved the attack success rate. Our attack algorithm has greatly improved its concealment and robustness.


In addition, we have also tried on real commercial voice equipment. We successfully tested the unmasked sound effect and the masked sound effect on the voice input of WeChat in the mobile phone and the voice keyboard of iFlytek, and used The high probability of adversarial samples to realize the control of such voice devices. However, in this type of real attack scenario, the impact of the environment and the playback device sometimes makes it impossible to complete an attack successfully at a time, and it often requires repeated playback of adversarial samples to achieve a targeted attack that is completely consistent with the target command. And for some instructions, there is still a problem of extremely poor attack efficiency.


It is worth mentioning that we tried a more demanding scenario, that is, in an office environment (with various types of noise), we used speakers to play the adversarial sample with masking music once and recorded it with a mobile phone. Then use the adversarial sample in the mobile phone to play and change to another mobile phone to record. After the second playback and recording, due to the different losses of different devices during recording and playback, the adversarial samples have lost more information and recorded significant environmental noise. From our perception, the quality of adversarial samples has been greatly reduced, and the adversarial attack can still be achieved with a certain probability, although it needs to increase the number of attempts. This experiment proved that the adversarial samples we generated have strong robustness and transmissibility, and they still maintain a certain attack effect after multiple passes. This experimental scenario has not been tried by previous adversarial attack algorithms.

\subsection{Evaluation on Hiding Effect}


When testing the masking effect, because everyone’s auditory perception is different, we evaluate the masking effect by inviting volunteers to do a questionnaire survey. Before the test, the volunteers were only told to listen to a piece of audio and give feedback on what they heard. Each volunteer will listen to the same audio twice (as the number of plays increases, volunteers will gradually realize that the audio contains additional information, and the probability of identifying hidden content will continue to increase). We divided the adversarial samples into two batches of unmasked sample and masked sample, then asked volunteers to randomly select from the two batches of samples for trial listening. During the experiment, in order to ensure the playback quality, the volunteers will wear the same headphones, and they can freely adjust the volume according to their needs.


It can be seen from the table \ref{tab:hiding} that in the first audition before adding the masking music, the volunteers did not recognize the hidden content, but some volunteers perceive that the audio contains information. In the second audition, three volunteers heard the instruction information contained in the adversarial sample. Most of the remaining volunteers only perceive the presence of noise and cannot confirm the content of the information. After the addition of masking music, no volunteers could perceive the instruction information, and a small number of volunteers felt that there was information, but they could not confirm the content of the information at all. Most volunteers only heard some ordinary instrument sounds when we asked, and did not feel that there was more information.

\begin{table}[]
\centering
\caption{Experimental results of hearing perception test after adding masking sound effect}
\begin{threeparttable}
\begin{tabular}{|c|cc|cc|}
\hline
          & \multicolumn{2}{c|}{Once-recognize}      & \multicolumn{2}{c|}{Twice-recognize}     \\ \hline
          & \multicolumn{1}{c|}{Volunteer \tnote{1}} & $\mathrm{Acc}$ \tnote{2}& \multicolumn{1}{c|}{Volunteer \tnote{1}} & $\mathrm{Acc}$ \tnote{2} \\ \hline
No Masking Music & \multicolumn{1}{c|}{0/10}   & 0    & \multicolumn{1}{c|}{3/10}   & 28.4 \\ \hline
Masking Music   & \multicolumn{1}{c|}{0/10}   & 0    & \multicolumn{1}{c|}{0/10}   & 0    \\ \hline
\end{tabular}
\label{tab:hiding}
\begin{tablenotes}
        \footnotesize
        \item[1] This indicator records (the number of volunteers who successfully identified hidden content)/(the total number of volunteers)
        \item[2] This indicator records the accuracy of volunteers' recognition of hidden content, that is, the ratio of the number of words correctly guessed to the total number of words.
      \end{tablenotes}
\end{threeparttable}
\end{table}



This experimental result shows that our adversarial attack algorithm has a certain degree of concealment when it is not equipped with masked music, but it is not strong enough. It is easy to guess the hidden content after repeated auditions. This is also a problem with most existing voice adversarial algorithms. After adding the masking music searched by the algorithm \ref{alg:search}, the concealment of the adversarial samples has increased significantly. According to the auditory experience of the volunteers, their attention is almost completely attracted by the masking music. Even if disturbing information is vaguely detected, it will be difficult to confirm the specific content under the influence of the masking music. This phenomenon is in line with the expected result of the masking effect of the psychoacoustic model.


For attackers, on the one hand, they have psychological expectations about the content of the instructions embedded in the audio, on the other hand, attackers who have heard a large number of adversarial samples naturally have some recognition skills. Therefore, attackers are usually unable to objectively evaluate the masking effect, and the masking effect will be weaker for them. In the actual attack, when the victim does not understand the form of the adversarial sample and has no recognition experience, it will be very difficult to identify the adversarial sample with masking music in a short period.

\subsection{Analysis}


\begin{figure*}[htbp]
\centering
\subfigure[Waveform of enter command read by human voice.]{
\includegraphics[width=6.0cm]{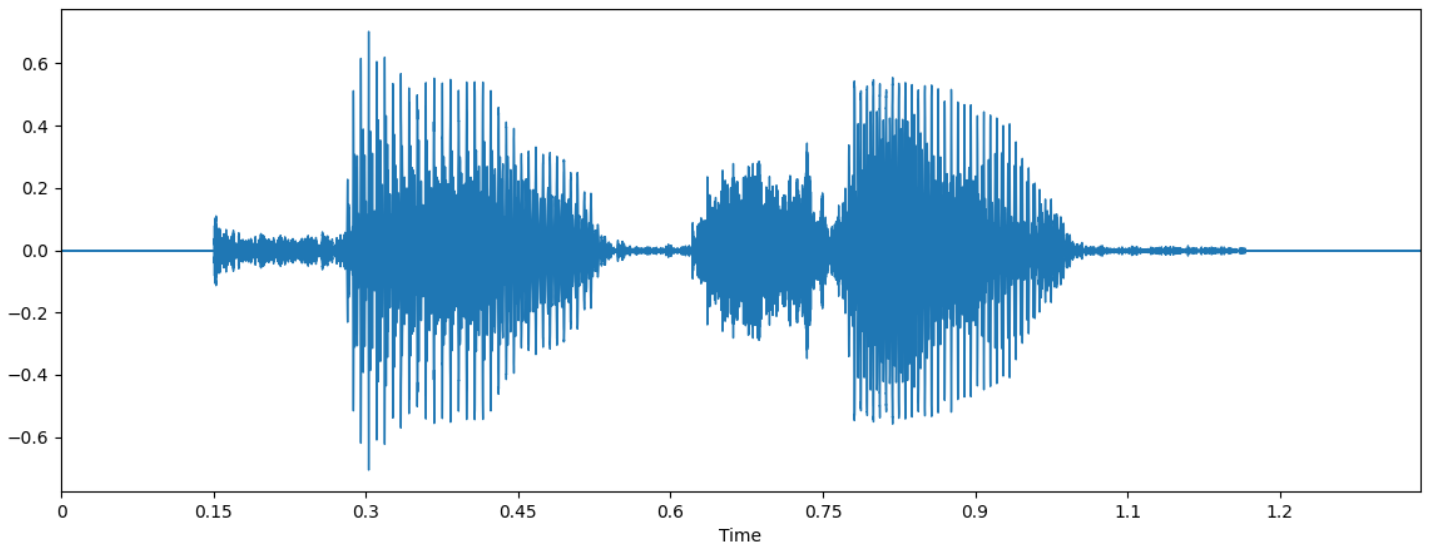}
}
\quad
\subfigure[Spectrogram of enter command read by human voice.]{
\includegraphics[width=6.0cm]{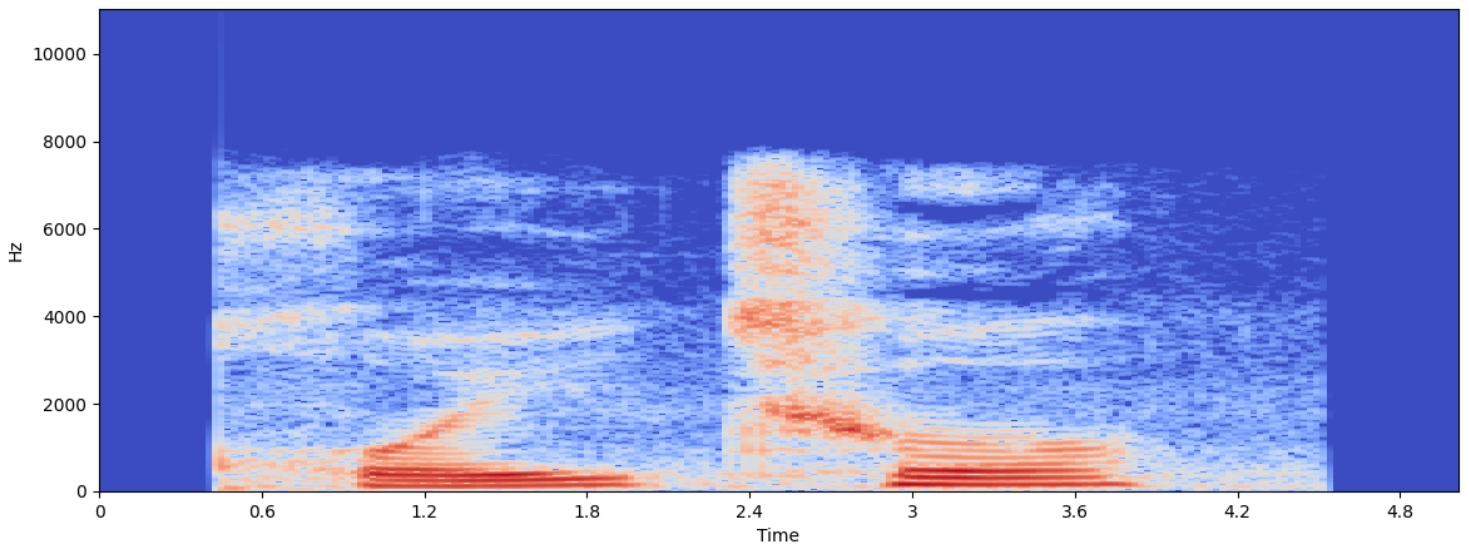}
}
\quad
\subfigure[Waveform of unmasked adversarial samples generated by the enter command.]{
\includegraphics[width=6.0cm]{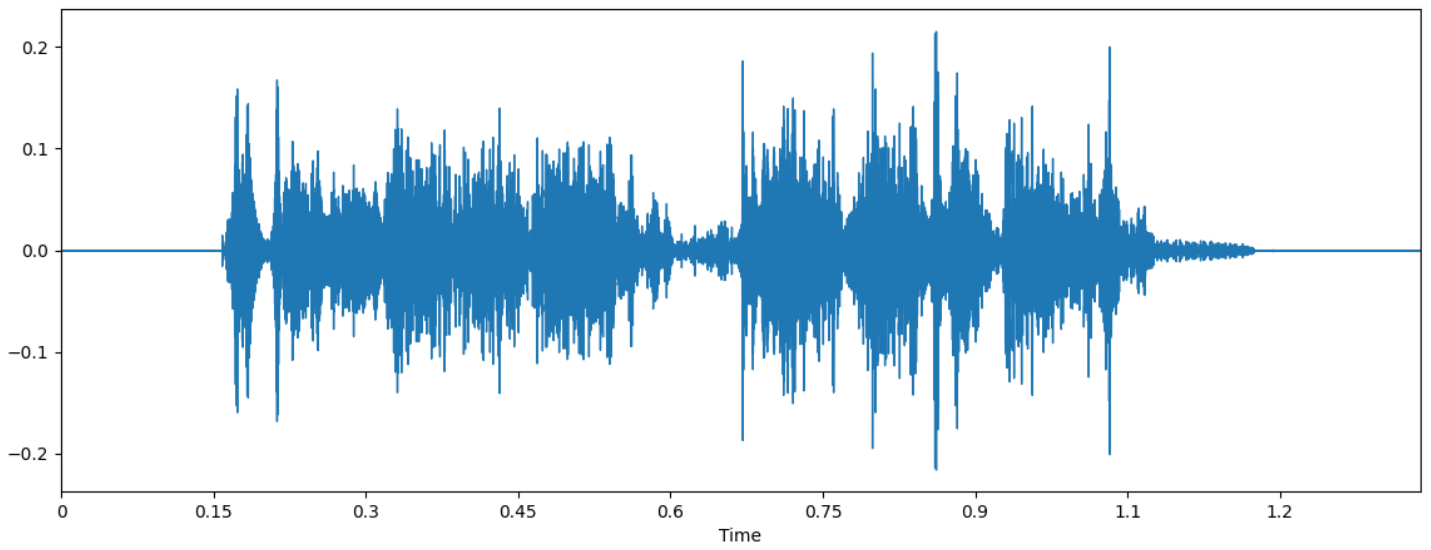}
}
\quad
\subfigure[Spectrogram of unmasked adversarial samples generated by the enter command.]{
\includegraphics[width=6.0cm]{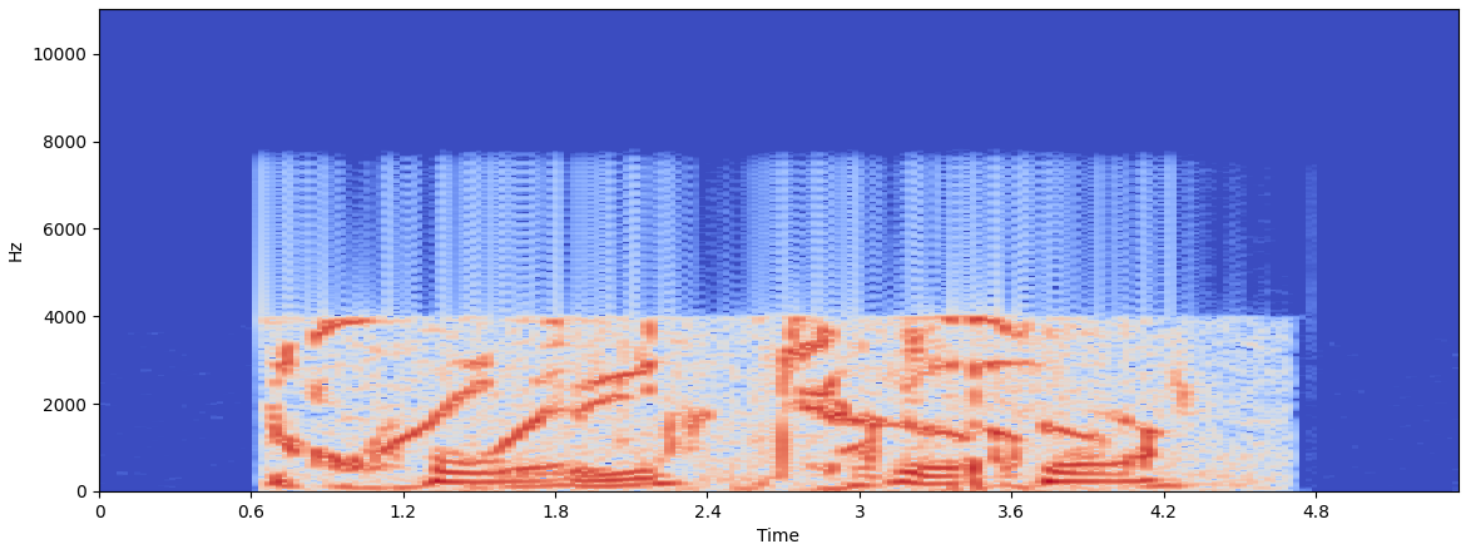}
}
\quad
\subfigure[Waveform of masking music generated for the adversarial sample of the enter command.]{
\includegraphics[width=6.0cm]{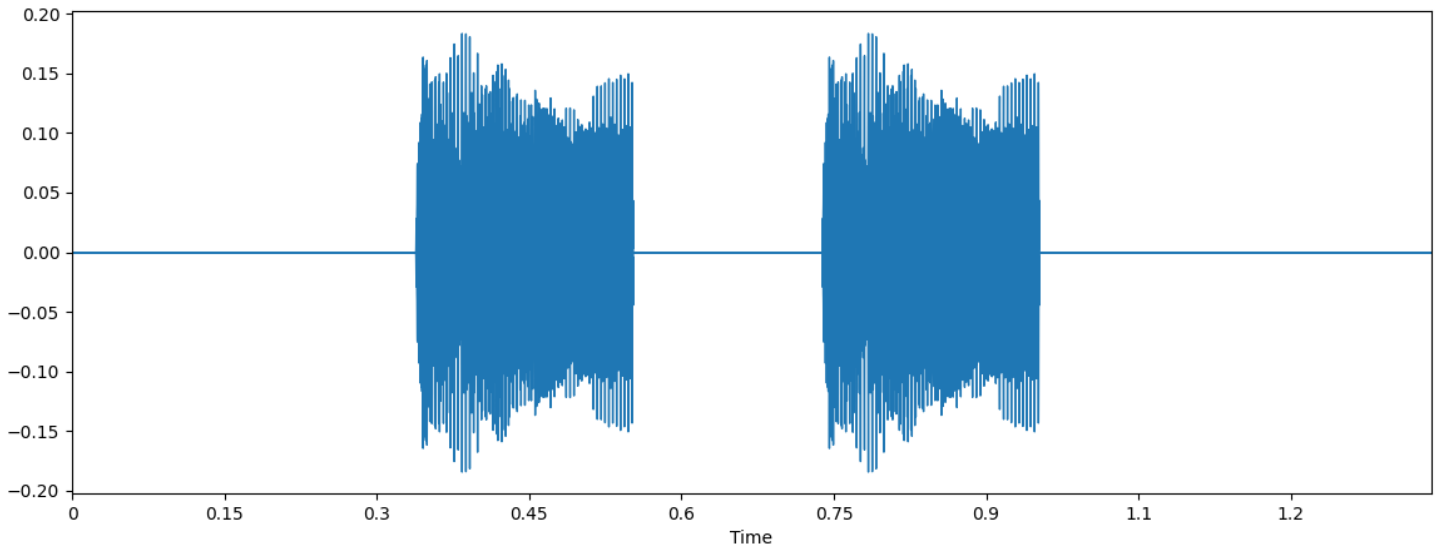}
}
\quad
\subfigure[Spectrogram of masking music generated for the adversarial sample of the enter command.]{
\includegraphics[width=6.0cm]{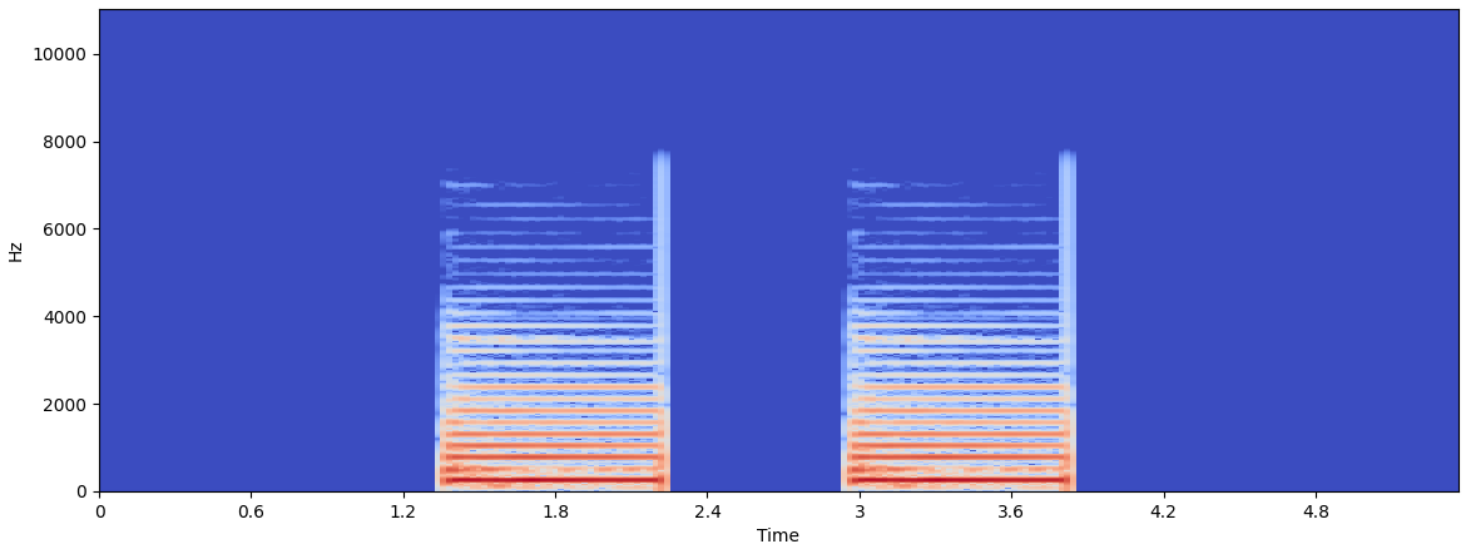}
}
\quad
\subfigure[Waveform diagram of the new adversarial sample synthesized from the adversarial sample of the enter command and the corresponding masking music.]{
\includegraphics[width=6.0cm]{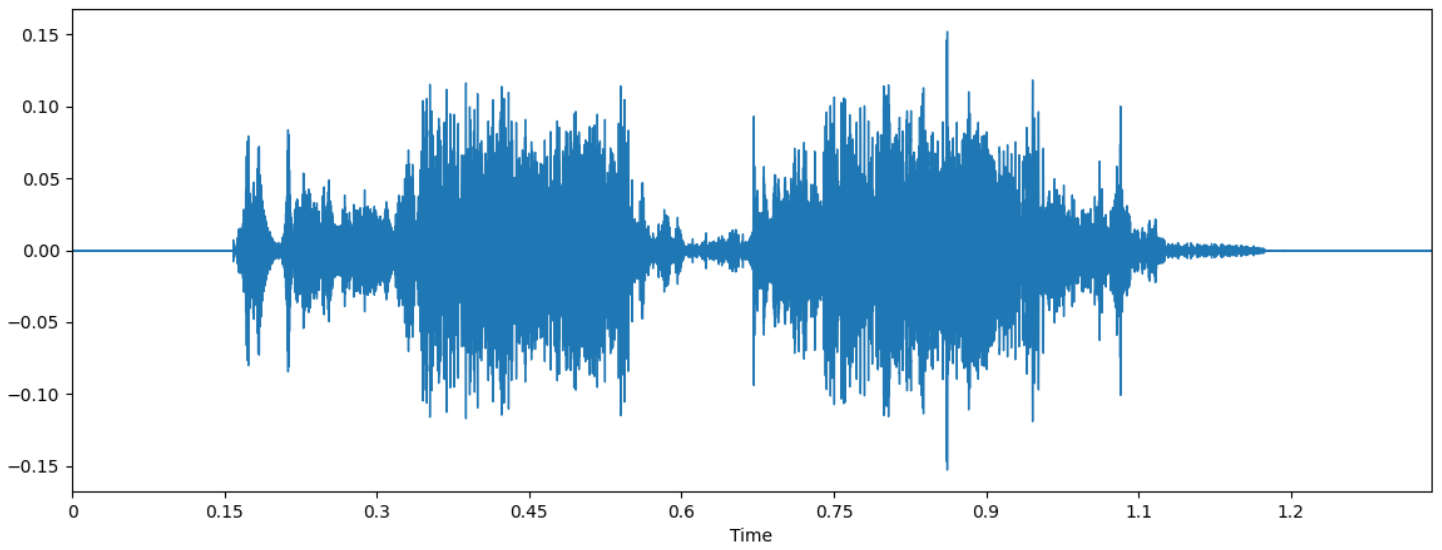}
}
\quad
\subfigure[Spectrogram of the new adversarial sample synthesized from the adversarial sample of the enter command and the corresponding masking music.]{
\includegraphics[width=6.0cm]{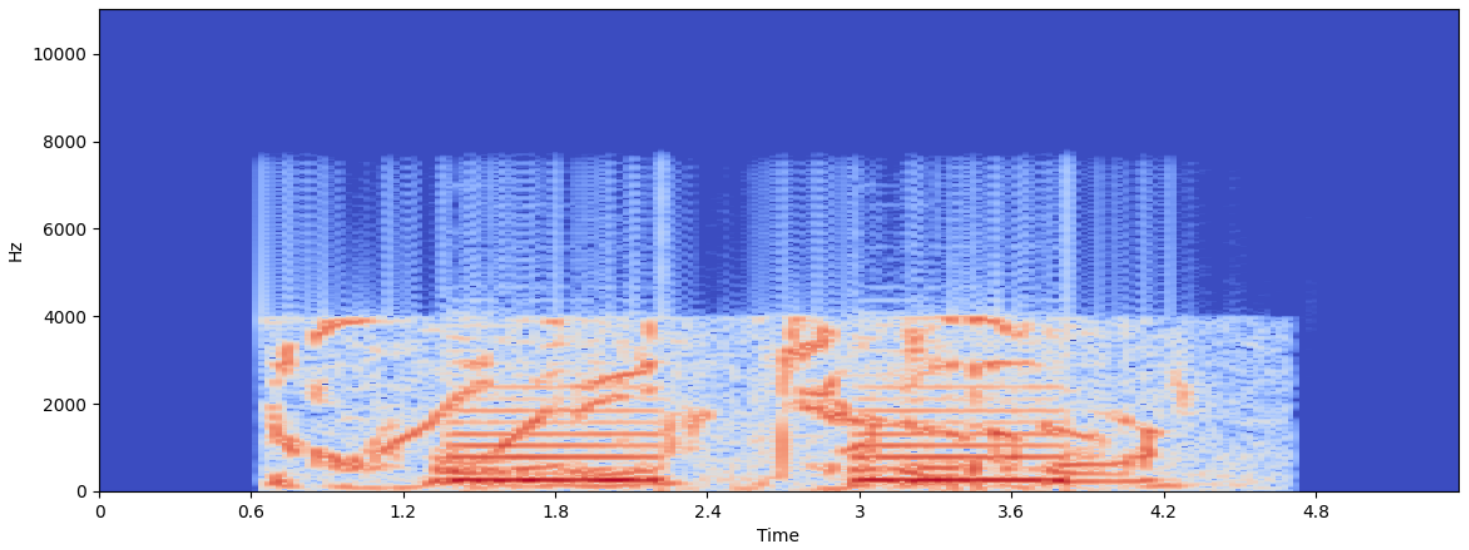}
}
\caption{Figures (a) and (b) are the waveform and spectrogram of the target command read by human voice (The voice content is "enter" command); Figures (c) and (d) are the waveform and spectrogram of adversarial samples without unmasked music; Figures (e) and (f) are the waveform and spectrograms of masked music searched based on the adversarial sample; Figures (g) and (h) are respectively the waveform and spectrogram of the adversarial sample after the masked music is added.}
\label{fig:res}
\end{figure*}


We selected the adversarial sample, the masking music, and the masking adversarial samples generated by the enter (Chinese pinyin is "hui che") command for analysis. The subgraph (a) and (b) in Figure \ref{fig:res} are waveform diagrams and spectrograms corresponding to the target instructions recorded by the human voice. From the comparison of (b) and (d) in Figure \ref{fig:res}, it is obvious that the confrontation samples of unmasked music retain a large number of low-frequency features consistent with the target instruction, although the corresponding waveforms are significantly different. These common features are important information for the ASR system to accurately identify target instructions and are also a guarantee for the high transferability of adversarial samples. The adversarial algorithm retains these features while also generating many messy high-frequency sounds. These features will make the adversarial sample sound more like noise and increase concealment.


The masking music waveform and spectrogram searched out based on Algorithm \ref{alg:search} are subgraph (e) and (f) in Figure \ref{fig:res}, respectively. Analyze from the time masking effect. The place that meets the simultaneous masking is the best masking position in the search results, which basically covers the most important features of the target instruction. In the actual audition, the place where the front of the waveform is not covered is not the main part of the pronunciation, so the masking sound is mainly in the second half of the pronunciation stage. From the analysis of the frequency domain masking effect, the best masking sound material searched in our experiment is the piano sound effect. The frequency of the masking sound material is similar to the frequency of the area where the key feature of the command is located, and it is easy to mask the original command sound. Only from the masking tone analysis found out, the frequency of the piano tone corresponding to the critical band on the Barker scale has larger bandwidth, and the masking effect is very strong. If an attacker wants to quickly make masking music for human voices, he can consider directly selecting piano sounds to mask key periods.


The subgraph (g) and (f) in Figure \ref{fig:res} are the waveform and spectrogram after the masking music and the adversarial samples are synthesized. From the waveform diagram, the synthesized sample is obviously different from the confrontation sample of unmasked music. But the effect from the spectrogram is similar to drawing a few horizontal lines on the picture. From the experience of confrontation attacks in the field of image recognition, although such a line may cause model misjudgment, it does not introduce other phoneme characteristics, and the introduced disturbance is not a gradient information-based confrontation disturbance generated by the algorithm. , Therefore, it is not easy to cause the effect of the adversarial attack and affect the migration of the adversarial sample itself. As for the ASR system that converts time-domain signals into frequency-domain signals for feature extraction, the audio recognition problem has been transformed into an image recognition problem. Therefore, although masking music is more noticeable in human auditory perception, it only occupies a small part of the characteristics in the frequency domain signal. This is also the reason why our masking music can improve the masking effect while having less impact on migration attacks.

The subgraph (g) and (f) in Figure \ref{fig:res} are the waveform and spectrogram after the masking music and the adversarial samples are synthesized. Although the introduction of additional masking music on the waveform diagram seems to have brought a greater change to the waveform, the impact on the spectrogram is similar to drawing a few horizontal lines on the picture. From the experience of adversarial attacks in the field of image recognition \cite{su2019one}, although such scribing may cause model misjudgment with a certain probability, it does not introduce other phoneme characteristics, and the introduced disturbance is not a gradient information-based disturbance generated by the adversarial algorithm, so it is not easy to cause the effect of adversarial attacks and affect the transferability of adversarial samples themselves. As for the ASR system that converts time-domain signals into frequency-domain signals for feature extraction, the audio recognition has been transformed into image recognition. Therefore, although masking music is more noticeable in human auditory perception, it only occupies a small part of the characteristics of the frequency domain signal. This is also the reason why our masking music can improve the masking effect while having less impact on transfer attacks.

\section{Discussion}

\subsection{Limitations}


In the process of confronting the attack, we will save the generated confrontation samples every few iterations. During the exploration process, we found that some instructions are efficient and stable in the confrontation attack. The migration attack can be realized when the number of iterations is not high, and the result of the migration attack changes little in the subsequent iteration process. Such samples are more robust against samples and have more stable and better effects when adding disturbance sounds or physical attacks. They almost all return similar results on different business models. 

On the contrary, some adversarial examples are not robust. The results they return on different commercial APIs are quite different, and they are susceptible to masking sounds and environmental factors in the physical world. This may be because the local model's data set has a small amount of training on these sentences, resulting in low quality of generation, or it may be that the homophones and synonyms of these sentences in the Chinese language environment increase the difficulty of fighting attacks. The low success rate is mainly due to the contributions of these samples.


As for concealment, by searching for masking music for adversarial samples, we have greatly improved the masking of adversarial samples. Volunteers who participated in the test did not understand the voice adversarial samples, nor did they anticipate what type of content the audio contained. Therefore, they were unable to identify adversarial samples with masked music at the first contact. However, as the number of plays increases, volunteers will gradually become aware of the information hidden in the audio, and volunteers with sensitive hearing will recognize adversarial samples faster. For researchers who are often exposed to voice adversarial samples, identifying hidden information is not too difficult. At present, the best-hidden work still needs to be generated on the white-box model. In addition, our masking sounds can only be pure music without semantic features at present, and cannot use the audio containing other sentences as a cover under the constraints of black-box attacks. The semantic features contained in this type of masking sound will greatly undermine the 
transferability of disturbance features.


In terms of robustness, as described in section \ref{sec:phy}, when the physical world is attacked, we use ordinary home audio and mobile phones as the playback and recording equipment. The experimental environment did not choose an empty and quiet room but an ordinary bedroom environment. Too many variables cause the same adversarial sample to be sometimes identified as different results during the experiment. In most cases, these results are similar. Because there are too many real environment variables, we cannot determine which variables are more likely to affect the success rate of attacks. For attack methods such as broadcast or Internet video propagation, there is still no guarantee of a 100\% success rate of targeted attacks.


In our experimental setting, in order to verify the triggering of malicious functions, the target instructions are usually 2-4 words, which are short instructions. When the attack requires that the target instruction is totally the same as the attack recognition result, the difficulty of the attack will increase as the length of the sentence increases. The black-box targeted attacks of long sentences are still not well resolved, which requires stable and robust adversarial attacks. In addition, for the popular mixed language model (that is, a language model can recognize audio that contains multiple languages), we still can't achieve a general attack, and can only use the corresponding language model to attack.


We have verified the transmissibility of adversarial samples in our experiments, but we have not conducted a deeper exploration of this phenomenon. Adversarial samples that can maintain high attack efficiency after repeated playback and recording will cause greater harm. We can construct such an attack scenario. The attacker spreads adversarial samples embedded with control instructions on the video platform. When the victim uses the speaker to play, the smart devices around him may be called, and the video is constantly being forwarded and disseminated. After the video is continuously forwarded and spread, the attack effect can still be maintained.

\subsection{Possible Defense}


In response to our proposed adversarial attack algorithm, we discuss several potential defense mechanisms.


\textbf{Audio Downsampling}. Both CommanderSong \cite{yuan2018commandersong} and Devil's Whisper attack \cite{chen2020devil} mentioned audio downsampling defenses. That is, audio can be recorded in different formats (such as m4a, mp3, wav) at different sampling rates (such as 8000Hz, 16000Hz, 48000Hz). For example, for an adversarial sample that can carry out adversarial attacking Tencent API, we first downsample it to 5000Hz, and then upsample it to 8000Hz, the success rate of the adversarial attack will be greatly reduced. In contrast, conventionally recorded human voices and TTS audio clips can still be correctly identified even after such downsampling or upsampling. In the process of multiple sampling, audio information will continue to be lost. Because the adversarial sample contains fewer voice features related to the target instruction than the actual target instruction sentence. Therefore, as the data quality declines, the attack efficiency will continue to decrease. 


\textbf{Adversarial Training}. Adversarial training was first proposed by Goodfellow\cite{goodfellow2014explaining}. The idea is simple and direct. The generated adversarial samples are added to the training set for training, and the robustness of the model is improved by data enhancement. Aleksander Madry et al.\cite{madry2017towards} used a stronger PGD adversarial attack algorithm to generate adversarial samples for adversarial training and further strengthened the robustness of the model. The disadvantage of adversarial training is that the training speed is slow and it may reduce the accuracy of the model's recognition of normal samples. Therefore, developers need to make a trade-off between robustness and accuracy.


\textbf{Adversarial Sample Detection}. Adversarial sample detection defends from the sample level, distinguishing and intercepting adversarial samples before entering the model. In our experiment, the adversarial sample is quite different from the normal sample. One way is to distinguish the samples by adding noise \cite{dong2021adversarial}. Adversarial samples are more susceptible to noise than normal sounds. In our experiment, after adding random white noise to the adversarial sample, as the noise volume increases, the attack success rate of the adversarial sample decreases much faster than the normal sample. From the auditory effect, the adversarial sample is also significantly different from the normal sample, and the audio will be accompanied by a clear sense of the noise. Therefore, training a classifier to recognize adversarial samples is also a potentially effective defense method.


\textbf{Speaker Recognition}. Adding speaker recognition to the ASR system is also a powerful and effective defense measure. The audio adversarial sample usually only retains the characteristics of the target instruction in the optimization process, and hardly contains the timbre characteristics of anyone. Most of the adversarial samples sounded noisy, closer to the timbre of the noise, and clearly distinguished from normal human speech audio. So the simple way is to detect whether the audio input is a human voice. A more secure way is to conduct stricter identity authentication for voice service users. By registering normal service user timbre as identity authentication, malicious invocation of voice services by others can be effectively avoided. AI fraud technologies such as Deepfake may bypass speaker authentication, but such attack methods are not easy to implement covertly.

\subsection{Future Work}


In subsequent experiments, we will try to increase the success rate of attacks against such sentences by adding corpus to the training set, especially for sentences that are difficult to attack. In addition, we will also try to see if there are any defects in the language model, that is, there are natural differences in the difficulty of attacking different sentences. On the other hand, we will also try to combine adversarial attacks with more application scenarios to discover new potential hazards. For example, adversarial samples are used to induce the navigation system to enter the wrong route. Adversarial attacks against special security scenarios will face more complex target sentences and real environments. We will also try some interesting attack schemes, such as adjusting the position of the voice device or using multiple devices to play different audio to achieve joint adversarial attacks. High transferability adversarial attacks are also one of the directions we want to continue to explore, which is related to the range of possible impacts of a malicious audio attack.


With the help of research experience in attacks, we will also explore more defense methods against attacks to help improve the security of AI models. As the application of AI voice systems in various scenarios increases, potential security risks will continue to increase. Voice adversarial attacks need to arouse more attention from AI voice researchers.

\section{Conclusion}


We propose an adversarial attack algorithm, which can achieve a success rate of 81.57\% without contacting the target model. It is currently the adversarial algorithm with the highest success rate among black box contactless adversarial attacks. We also propose a masked music search algorithm based on psychoacoustics to protect adversarial samples. This is the first exploratory work on adapting masking music to a fixed adversarial sample. Under our masked music, no volunteers can identify the content contained in adversarial samples in experiments. The two algorithms complement each other and take into account the transfer success rate and concealment of the adversarial attack. Our experiments are all carried out on commercial speech recognition systems, and we have simulated the environment and possible equipment in real attack scenarios as much as possible during the experiments in the physical world. The attack algorithm we proposed shows that the current ASR system is still not robust and secure. As the ASR system becomes more and more popular in applications, the security problems faced will become increasingly serious. We hope that our research can arouse the attention of AI researchers and developers to safety\cite{2204.08977}.





%

\bibliographystyle{ieeetr}
\bibliography{ref}

\end{document}